\documentclass[%
preprint,
superscriptaddress,
showpacs,
showpacs,
preprintnumbers,
amsmath,
amssymb,
aps,
]{revtex4-1}

\usepackage[dvipdfmx]{graphicx}
\usepackage{dcolumn}
\usepackage{bm}
\usepackage{multirow}
\usepackage[mathlines]{lineno}
\usepackage{comment}

\usepackage{xr}

\begin{document}

\preprint{}

\title{
Realization of nodal ring semimetal in pressurized black phosphorus
}

\author{Kazuto~Akiba}
\email{akb@okayama-u.ac.jp}
\affiliation{
Graduate School of Natural Science and Technology,
Okayama University, Okayama 700-8530, Japan
}

\author{Yuichi~Akahama}
\affiliation{
Graduate School of Science, University of Hyogo, Kamigori, Hyogo 678-1297, Japan
}

\author{Masashi~Tokunaga}
\affiliation{
The Institute for Solid State Physics, The University of Tokyo,
Chiba 277-8581, Japan
}

\author{Tatsuo~C.~Kobayashi}
\affiliation{
Graduate School of Natural Science and Technology,
Okayama University, Okayama 700-8530, Japan
}

\date{\today}

\begin{abstract}
Topological semimetals are intriguing targets for exploring unconventional physical properties of massless fermions.
Among them, nodal line or nodal ring semimetals have attracted attention
for their unique one-dimensional band contact in momentum space and resulting nontrivial quantum phenomena.
By field angular resolved magnetotransport measurements and theoretical calculations, we show that pressurized black phosphorus (BP) is an ideal nodal ring semimetal with weak spin-orbit coupling, which has a sole and carrier density-tunable nodal ring isolated from other trivial bands.
We also revealed that the large magnetoresistance effect and its field-angular dependence in semimetallic BP
are due to highly anisotropic relaxation time.
Our results establish pressurized BP as an elemental model material for exploring nontrivial quantum properties unique to the topological nodal ring.
\end{abstract}

\maketitle

Three-dimensional topological semimetals (TSMs) are a major research topic at the forefront of condensed matter physics \cite{Lv_2021}.
A notable feature of TSMs is massless fermions in the bulk, which derive from the linearly dispersing band structure.
Their unconventional physical properties due to the presence of massless fermions have been intensively explored and elucidated.
Dirac and Weyl semimetals \cite{Burkov_WS,Wang_2013,Liu_Na3Bi,Liu_Cd3As2,Xu_TaAs,Wu_WTe2,Deng_MoTe2} are widely known materials that are characterized by degenerate or non-degenerate band contact points in the momentum space.
Another type of TSM is a nodal line or nodal ring semimetal \cite{Burkov_NS},
in which the contact of the valence and conduction band occurs on a continuous line or closed ring in the momentum space.
Numerous unique physical properties characterizing the nodal line or nodal ring semimetal are expected, including a drumhead-like flat surface state \cite{Kim_2015},
unconventional distribution of Landau levels \cite{Rhim_2015},
nontrivial $\pi$-Berry phase associated with the band contact line \cite{Li_2018} etc.

However, an ideal TSM is rarely seen in reality; other trivial carriers generally co-exist other than the massless fermions, which hinder the extraction of the true physical response derived from the topological electronic structure.
In addition, the topological character is inherent in the crystal structure, constituent element, etc.
which are difficult to desirably control from the outside.

\begin{figure}[]
\centering
\includegraphics[]{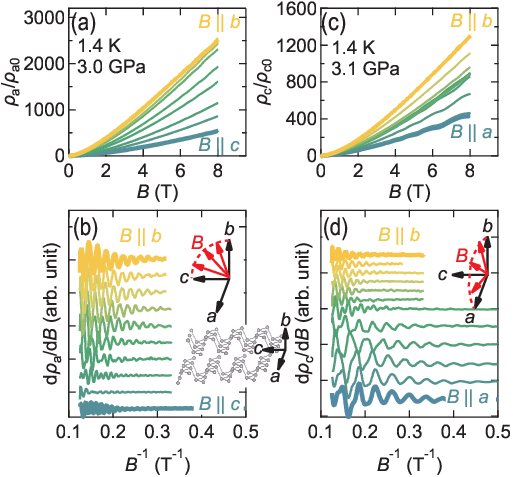}
\caption{
(a) Magnetic field dependence of resistivity $\rho_{a}$ normalized by zero-field value $\rho_{a0}$ and (b) $d\rho_{a}/dB$ at 1.4 K and 3.0 GPa.
$B$ is rotated within the $bc$ plane, and $I$ is along the $a$ axis.
(c) Magnetic field dependence of resistivity $\rho_{c}$ normalized by zero-field value $\rho_{c0}$ and (d) $d\rho_{c}/dB$  at 1.4 K and 3.1 GPa.
$B$ is rotated within the $ab$ plane, and $I$ is along the $c$ axis.
Data in (b) and (d) are vertically shifted for clarity.
The crystal structure of BP and its crystal axes are shown in the insets of (b) and (d).
\label{fig01}}
\end{figure}

In this context, we focus on black phosphorus (BP),
which is known as a monoatomic semiconductor with a band gap of 0.3 eV at the $Z$ point in the first Brillouin zone at ambient pressure \cite{Takao_1981, Asahina_1982, Akahama_1983}.
The crystal structure has puckered honeycomb layers, which are alternately stacked along the $b$ axis, as shown in the inset of Fig. \ref{fig01}(b)  \cite{Hultgren_1935, Brown_1965}.
BP shows a pressure-induced semiconductor-to-semimetal transition at $\sim1.4$ GPa
\cite{Akahama_1986, Akahama_2001, Akiba_2015, Xiang_2015, Fujii_2020}.
In the semimetallic state, the carrier density of electrons and holes can be continuously tuned by applying pressure, maintaining its compensated nature and high carrier mobility \cite{Akiba_2017}.
There have been several suggestions for the electronic structure in the semimetallic state \cite{Akiba_2015, Fei_2015, Gong_2016, Li_2017,Fujii_2023},
including the possible realization of topological nodal line semimetal \cite{Zhao_2016}.
Intriguingly, several materials such as the nonmagnetic CaP$_3$ family \cite{Xu_2017, Kim_2023}
and magnetic EuP$_3$ \cite{Mayo_2022},
which share a similar puckered-layer structure with BP, have been focused on as candidates of nodal ring semimetal.
Experimentally, however, the Fermi surface (FS) in the semimetallic BP has been veiled so far primarily because of the lack of detailed geometry under high pressure.

In this study, we unveil the FS of semimetallic BP by angular resolved magneto transport
and show that pressurized BP is an ideal topological nodal ring semimetal.
Single crystals of BP were synthesized under high pressure \cite{Endo_1982, Akahama_2020},
which show a high residual resistivity ratio of more than 400 at 3.4 GPa, as shown in Supplemental Material \cite{SM_URL}.
The orthorhombic crystal structure ($Cmca$, space group \#64) and the crystal orientation were determined by X-ray Laue backscattering image and simulations using QLaue \cite{qlaue_url}.
Laue pattern (see Supplemental Material \cite{SM_URL}) was reasonably explained by the
lattice parameters determined by a recent neutron diffraction
measurement \cite{Akahama_2020}.
Resistivity measurements under high pressures up to 3.5 GPa were performed using an indenter-type pressure cell \cite{Kobayashi_2007} and Daphne oil 7474 pressure medium \cite{Murata_2008}.
The pressure in the sample space was determined from the superconducting transition temperature of Pb set near the sample \cite{Eiling_1981}.
The band structure calculations based on the density-functional theory (DFT)
and subsequent analyses 
were performed using the Quantum ESPRESSO \cite{Giannozzi_2009, Giannozzi_2017, Giannozzi_2020},
Wannier90 \cite{Pizzi_2020},
WannierTools \cite{Wu_2017, Zhang_2019},
FermiSurfer \cite{Kawamura_2019},
and the SKEAF code \cite{Rourke_2012}.
In the following, we discuss the results based on scalar-relativistic calculations since the effect of the spin-orbit coupling is negligibly small.
A more detailed description of the experimental and theoretical methods
are shown in Supplemental Material \cite{SM_URL}.


First, we focus on the magnetotransport properties and their field angular dependence in the semimetallic state.
Figure \ref{fig01}(a) shows the magnetic field ($B$) dependence of $\rho_{a}$ at 1.4 K and 3.0 GPa, which is normalized by the zero-field value $\rho_{a0}$.
Here, we rotated $B$ within the $bc$ plane.
The current ($I$) direction is perpendicular to $B$.
We observed a large magnetoresistance (MR) effect, which was maximized by the application of $B$ along the $b$ axis.
To extract the SdH oscillations on such a large MR, we show $d\rho_{a}/dB$ as a function of $B^{-1}$ in Fig. \ref{fig01}(b).
We can see clear SdH oscillations and their variations according to the field direction.
In Figs. \ref{fig01}(c, d),
we show the resistivity along the $c$ axis normalized by the zero-field value
($\rho_{c}/\rho_{c0}$) at 1.4 K and 3.1 GPa, in which $B$ was rotated within the $ab$ plane.
In this configuration, we also observed a large MR effect maximized by the application of $B\parallel b$,
and prominent SdH oscillations.

\begin{figure}[]
\centering
\includegraphics[]{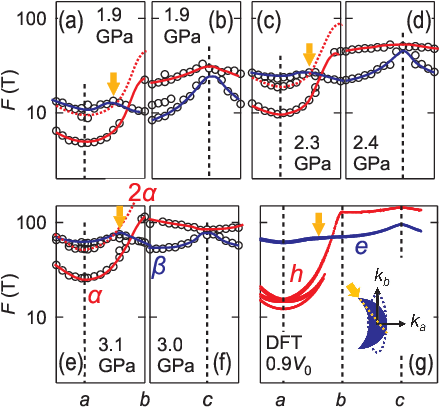}
\caption{
Field angular dependence of the SdH frequency $F$ at (a, b) 1.9 GPa, (c, d) 2.3-2.4 GPa, and (e, f) 3.0-3.1 GPa.
Identified branches ($\alpha$, $\beta$) and second harmonics ($2\alpha$) are shown with solid and broken eye guides, respectively.
(g) Field angular dependence of $F$ obtained from DFT calculation.
The inset in (g) represents possible deformation of the electron pocket from the perfect ellipsoid, which can cause local maxima between $a$ and $b$, as indicated by arrows.
\label{fig02}}
\end{figure}

By collecting the field angular resolved SdH oscillations at various pressures, we can obtain the detailed geometry and pressure-induced evolution of the FS.
Figures \ref{fig02}(a--f) show the field angular dependence of the SdH frequency ($F$) obtained by fast Fourier transform (FFT) analysis at three representative pressures.
The FFT spectra used to construct Figs. \ref{fig02}(a--f) are shown in Supplemental Material \cite{SM_URL}.
Although the datasets shown in (a, c, e) and (b, d, f) are obtained using different setups and samples,
we can observe reasonable reproducibility of $F$ at approximately the same pressure.
At all pressures, we identified two independent branches labeled $\alpha$ and $\beta$ with eye guides.
Here, we can see another frequency around $B\parallel a$,
which lies just below the $\beta$.
We confirmed at all pressures that these are second harmonics of $\alpha$.
The observation of $2\alpha$ is consistent with our previous study \cite{Akiba_2015}.
We can see that $F$ of $\alpha$ and $\beta$ monotonically increases with the application of pressure with little change in their angular dependencies.
The monotonic increase of $F$
corresponds to the increase in the carrier density indicated by the two-carrier model analyses \cite{Akiba_2017}.
Here, we can recognize a pressure-induced change in the angular dependence of $\alpha$ between $B\parallel b$ and $B\parallel c$, which will be discussed later.
From the above results, we can conclude there exist two independent Fermi pockets and no other FSs appear at least up to 3.0--3.1 GPa.


\begin{figure}[]
\centering
\includegraphics[]{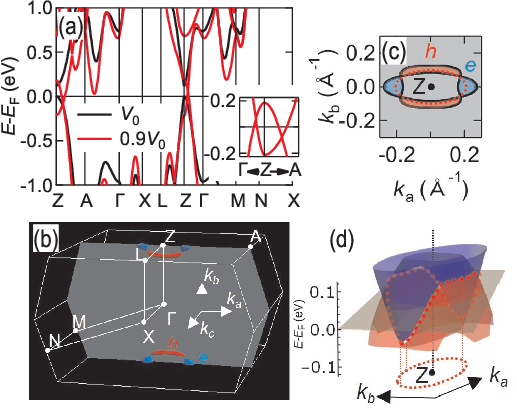}
\caption{
(a) Energy band structure under ambient (black) and compressed (red) conditions.
$V_0$ represents the volume of the optimized structure at ambient pressure (see text for details).
The inset shows a magnified view of the band dispersion at the $Z$ point.
(b) FS under compressed conditions.
The red and blue surfaces represent the hole and electron pockets, respectively.
(c) Cross-section of the FS cut by $k_a$--$k_b$ plane.
The red broken line indicates the projection of the nodal ring on the $k_a$--$k_b$ plane.
(d) Dispersion of the valence (red surface) and conduction (blue surface) bands on the $k_a$--$k_b$ plane.
Red broken lines represent the nodal ring enclosing the $Z$ point.
\label{fig03}}
\end{figure}

To determine the FS of the semimetallic BP, we performed first-principles calculations.
The black curve in Fig. \ref{fig03}(a) shows the band structure of BP at ambient pressure using a fully optimized structure, which reproduces the direct band gap structure at the $Z$ point.
Here, $V_0=171.0$ \AA$^3$ in Fig. \ref{fig03}(a) represents the volume of the unit cell obtained by structural optimization, and the Fermi level is taken at the top of the valence band.
The band structure at ambient pressure is consistent with a previous report \cite{Qiao_2014}.
The red curve in Fig. \ref{fig03}(a) shows the band structure of compressed BP with a volume of $0.9 V_0$.
The result was obtained using a crystal structure optimized under hydrostatic compression, and the Fermi level was adjusted to satisfy the carrier compensation.
For details of structural optimization, see Supplemental Material \cite{SM_URL}.
As seen at the $Z$ point, the valence and conduction band touch at a point near the Fermi level.
Although the conduction band significantly falls toward the Fermi level on the $\Gamma$--$A$ path, the present result indicates that the emergence of the FS initially occurs at the $Z$ point.
This picture is qualitatively different from those suggested in several previous reports \cite{Akiba_2015, Gong_2016},
in which four additional electron pockets exist on the $\Gamma$--$A$ path.
A possible reason for this difference might account for the computational details and whether the structure was optimized or not.

Figure \ref{fig03}(b) shows the FS drawn in the first Brillouin zone.
We obtained an elongated banana-shaped hole ($h$) pocket and a relatively isotropic electron ($e$) pocket.
As shown in Fig. \ref{fig03}(c),
these pockets touch at four nodes and enclose the $Z$ point in the $k_a$--$k_b$ plane.
To deepen our understanding of the obtained FS, we show in Fig. \ref{fig03}(d) the dispersion of the valence and conduction bands on the $k_a$--$k_b$ plane.
As indicated by the red broken line, the band-touching point mentioned in Fig. \ref{fig03}(a) forms a closed ring structure in the $k_a$--$k_b$ plane.
This nodal ring does not lie exactly on the Fermi level but shows a slight dispersion of $ \pm \sim100$ meV around the Fermi level.
This results in a squeezed ring-shaped FS formed by $h$ and $e$ pockets.
The above results predict the realization of a nodal ring semimetal,
in which there only exists a set of FSs deriving from the nodal ring structure.
The above discussion can hardly be affected by whether spin-orbit coupling is included or not, as shown in Supplemental Material \cite{SM_URL}.
This indicates that BP under pressure is a unique platform possessing an ideal nodal ring located within $\pm \sim100$ meV around the Fermi level.

Then, we confirm the realization of the predicted FS.
Figure \ref{fig02} (g) shows the field angular dependence of $F$ calculated based on the obtained FS.
As expected from the banana-shaped geometry,
$h$ pocket shows a steep angular dependence between $a$ and $b$.
The $e$ pocket shows rather flat dependence with local maximum and minimum at $c$ and $a$, respectively.
A comparison with Figs. \ref{fig02}(e) and (f) shows
that the experimental features in the angular dependence and absolute value of $F$
are satisfactorily reproduced by our calculation, demonstrating the realization of the nodal ring semimetal.

In both the experimental and theoretical results, we can recognize a local maxima between $a$ and $b$
indicated by the arrows in Fig. \ref{fig02}.
Qualitatively, these can be understood assuming that the poles of an ellipsoidal FS deviate from $k_b$, as illustrated in the inset of Fig. \ref{fig02}(g).
This structure corresponds to the pointy contacts directed along the nodal ring suggested by the DFT results shown in Figs. \ref{fig03}(b) and (c).
From the position of the local maxima, we determined that the tangent line of the nodal ring at the contact point of $h$ and $e$ pockets directs $45^\circ$--$55^\circ$ from $k_b$.

Here, we also note in Figs. \ref{fig02}(a--f) that the field angular dependence of $\alpha$
between $B\parallel b$ and $B\parallel c$ apparently changes by application of pressure.
At 1.9 GPa, $h$ pocket is assumed to be slightly flat on the $k_a$--$k_b$, which results in a larger cross-section when $B\parallel c$.
At 3.0-3.1 GPa, on the other hand, the FS becomes flat on the $k_c$--$k_a$ plane
as can be seen in the larger $F$ when $B\parallel b$.
We show the schematic images of the expected geometry in Supplemental Material \cite{SM_URL}
to help understand the situation.
The above change of flat direction is assumed to continuously take place by the application of pressure.
This delicate change resolved in our experiment could not be reproduced by the simulation presumably due to the potential difficulty to evaluate a tiny Fermi surface with high accuracy in the first-principles calculations.

\begin{figure}[]
\centering
\includegraphics[]{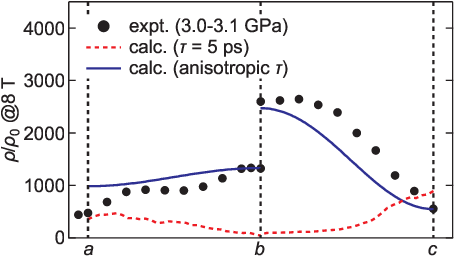}
\caption{
(a) Field angular dependence of the MR effect at 8 T with $I \perp B$.
The markers represent the experimental results.
The red broken and blue solid lines represent the calculation results assuming a constant relaxation time $\tau =5$ ps and anisotropic $\tau$, respectively (see main text).
\label{fig04}}
\end{figure}

Next, we attempt to understand the large MR effect characterizing the semimetallic state.
First, we focus on the transport properties arising from the FS curvature.
We theoretically evaluated the MR effect based on the semiclassical Boltzmann equation, whose result is shown by the red broken line in Fig. \ref{fig04} (for details, see Supplemental Material \cite{SM_URL}).
Here, we assume that the relaxation time $\tau = 5$ ps is independent of the band indices and directions.
Apparently, the calculation fails to reproduce the magnitude and field angular dependence of the MR observed in the experiments.

This mismatch is assumed to be derived from the constant-$\tau$ approximation.
To consider the possible band-index dependence and anisotropy of $\tau$,
we adopt a model
in which the mobility tensor of $h$ and $e$ pockets $\hat{\mu}^{h, e}$ and the magnetic tensor $\hat{B}$ are given by \cite{Mitani_2020}
\begin{equation}
\hat{\mu}^{h, e}=e
\begin{pmatrix}
\tau_{xx}^{h,e}/m_{xx}^{h,e} & 0 & 0 \\
0 & \tau_{yy}^{h,e}/m_{yy}^{h,e} &0 \\
0 &0 & \tau_{zz}^{h,e}/m_{zz}^{h,e}
\end{pmatrix},
\end{equation}
\begin{equation}
\hat{B}=
\begin{pmatrix}
0 & -B_z & B_y \\
B_z & 0 & -B_x \\
B_y & B_x & 0
\end{pmatrix}.
\end{equation}
Here, $x$, $y$, and $z$ correspond to $a$, $c$, and $b$ directions, respectively.
In this framework, the geometry of the FS is approximated to an ellipsoid characterized by the effective mass tensor $\hat{m}^{h, e}$.
On the other hand, we can treat $\tau$ as a pocket- and direction-dependent tensor $\hat{\tau}^{h, e}$.
The conductivity tensor $\hat{\sigma}$ and resistivity tensor $\hat{\rho}$ can be given by
$\hat{\sigma}=2[n_h e (\hat{\mu}^{h}+\hat{B})^{-1}+n_e e (\hat{\mu}^{e}-\hat{B})^{-1}]$ and
$\hat{\rho}=\hat{\sigma}^{-1}$.
To reproduce the experimental data at 3.0-3.1 GPa shown by the markers in Fig. \ref{fig04},
we calculated the resistivity normalized by the zero-field value by adjusting the six parameters in $\hat{\mu}^{h, e}$.
A reasonable reproduction was achieved as indicated by the solid lines in Fig. \ref{fig04},
in which $\mu_{xx}^{h}=1.5$ T$^{-1}$, $\mu_{yy}^{h}=10$ T$^{-1}$, $\mu_{zz}^{h}=12$ T$^{-1}$ for $h$ and
$\mu_{xx}^{e}=8$ T$^{-1}$, $\mu_{yy}^{e}=8$ T$^{-1}$, $\mu_{zz}^{e}=1$ T$^{-1}$ for $e$.
Here, we assumed a slight carrier imbalance,
$n_h=3.05\times10^{18}$ cm$^{-3}$ and $n_e=3.00\times10^{18}$ cm$^{-3}$ for each pocket, to reflect the positive Hall resistivity in the high-field region \cite{Akiba_2017}.
In fact, this model reasonably reproduces the experimental features of Hall resistivity, including the sign inversion in the low-field region (see Supplemental Material \cite{SM_URL}).
Thus, the above parameters depict a minimal model for expressing the transport properties of semimetallic BP.

Since the geometry of the FS has been established by first-principles calculations,
we can calculate the cyclotron effective mass $m_c$ of each pocket when $B$ is applied along the three principal axes.
The results are $m_c^{B\parallel a}/m_0=0.0476$, $m_c^{B\parallel b}/m_0=0.302$, $m_c^{B\parallel c}/m_0=0.318$ for $h$,
and
$m_c^{B\parallel a}/m_0=0.106$, $m_c^{B\parallel b}/m_0=0.151$, $m_c^{B\parallel c}/m_0=0.171$ for $e$,
where $m_0$ represents the bare mass of electron.
These show reasonable agreement with the experimental value,
$m_c/m_0=0.02$--0.14 \cite{Akiba_2015, Xiang_2015}.
Thus, we can evaluate $\hat{m}^{h, e}$ as:
$m_{xx}^{h}/m_0=2.02$, $m_{yy}^{h}/m_0=0.0453$, $m_{zz}^{h}/m_0=0.0500$ for $h$,
and
$m_{xx}^{h}/m_0=0.244$, $m_{yy}^{h}/m_0=0.0936$, $m_{zz}^{h}/m_0=0.121$ for $e$.
Here, we assumed relationships $m_c^{B\parallel x}=\sqrt{m_{yy} m_{zz}}$ etc.
Combining the above $\hat{m}^{h, e}$ with the experimentally deduced $\hat{\mu}^{h, e}$,
we can extract the relaxation time in ps as
\begin{equation}
\hat{\tau}^{h}=
\begin{pmatrix}
17.2 & 0 & 0 \\
0 & 2.58 &0 \\
0 &0 & 3.41
\end{pmatrix},
\hat{\tau}^{e}=
\begin{pmatrix}
11.1 & 0 & 0 \\
0 & 4.26 &0 \\
0 &0 & 0.685
\end{pmatrix}.
\end{equation}
The above result reveals that $\tau$ strongly depends on the direction and type of pocket, which is the reason for the failure of the constant-$\tau$ approximation.
Notably, anisotropic $\hat{\mu}^e$ is crucial to explain the large MR effect and its field angular dependence.
Because the relatively isotropic geometry of the $e$ pocket is validated by both experiments and calculations, the significant anisotropy of $\hat{\mu}^e$ is primarily responsible for $\hat{\tau}^e$.

Phonon is almost inactive at 1.4 K; thus, carrier-defect or carrier-carrier scattering may be responsible for the anisotropic $\hat{\tau}^{h,e}$.
Although the specific mechanism is unclear at this stage, we would like to comment on possible causes.

One is the anisotropic crystal structure of BP.
The $a$ direction having the longest $\tau$ for both $h$ and $e$ corresponds to the direction along the grooves of the puckered layer.
Thus, it might be preferable for ballistic transport compared to the buckled and layered directions.
We also note that the compressibility of BP is strongly direction-dependent.
As shown in a previous report \cite{Akahama_2020},
the lattice constant along the $a$ axis is hardly affected by the application of pressure, implying fewer defects accompanied by lattice compression.

Another one is unconventional scattering process, which is theoretically predicted
in nodal ring semimetal.
In particular, FS accompanied by a nodal ring can bring a two-dimensional weak antilocalization (2D WAL) effect
assuming a long-range impurity potential \cite{Syzranov_2017, Chen_2019}.
In this case, the backscattering process, which dictates the electrical resistivity,
is dominated by an interference loop confined in a specific plane perpendicular to the nodal line.
The longer $\hat{\tau}^{h,e}_{xx}$ may involve protection from backscattering under the presence of the 2D WAL effect.
It has been suggested that
a high aspect ratio between the radius of the nodal ring ($K_0$) and the poloidal radius of the FS ($\kappa$) in momentum space
is favorable for the emergence of the 2D WAL effect \cite{Chen_2019, Kim_2023}.
Based on $K_0\sim 0.15$ \r{A}$^{-1}$
and $\kappa\sim 0.036$ \r{A}$^{-1}$
obtained from the computational result shown in Fig. \ref{fig03}, $K_0/\kappa\sim 4.2$ can be estimated for BP at around 3 GPa.
Here, $K_0$ is estimated as an averaged value of the long and short axes of the oval nodal ring,
and  $\kappa$ is calculated using an average of extremal cross-sections of $h$ and $e$ pockets.
We would like to point out that this is comparable to the case of SrAs$_3$ ($K_0/\kappa\sim$ 3.3--4.5) \cite{Kim_2023},
which has a torus-shaped FS with a high $K_0/\kappa$ ratio and shows a 2D WAL effect.
To reinforce the above nontrivial mechanism,
further detailed and systematic inspection of electrical conductivity in the semimetallic state would be necessary,
which remains a future challenge.

In summary, we have unveiled the Fermi surface (FS)
in semimetallic black phosphorus (BP) by magnetotransport measurements.
We identified two independent FSs in the semimetallic state above 1.4 GPa, and no other FSs were detected up to 3.0--3.1 GPa.
Our theoretical calculation suggested the emergence of the FS at the $Z$ point in the first Brillouin zone and the realization of the nodal ring structure enclosing the $Z$ point.
The nodal ring has an energy dispersion of $ \pm \sim100$ meV around the Fermi level, which results in a squeezed ring-shaped FS consisting of small hole and electron pockets.
Because of the weak spin-orbit coupling of phosphorus, the gap at the band contact node is quite small.
The simulated angular dependence of the Shubnikov-de Haas oscillations satisfactorily reproduced the experimental features, which clarified the realization of the nodal ring semimetal.
We also demonstrated that anisotropic relaxation time is crucial for reproducing the large magnetoresistance effect and its field angular dependence in semimetallic BP.
Importantly, the semimetallic BP has the sole nodal ring isolated from other trivial bands, and the figure of the nodal ring and carrier density can be flexibly tuned by pressure.
Our results have established the realization of an ideal nodal ring semimetal in pressurized BP, which can be a promising platform to systematically explore novel phenomena derived from the nontrivial electronic structure.

\begin{acknowledgments}
We thank Y. Fuseya, H. Sakai, Y. Seo, and S. Araki for the valuable discussion.
This study was supported by JSPS KAKENHI Grants No. 22K14006 and 23H04862.
\end{acknowledgments}


\bibliography{reference}

\end{document}


\setcounter{equation}{0}
\setcounter{figure}{0}
\setcounter{table}{0}
\setcounter{page}{1}
\makeatletter
\renewcommand{\theequation}{S\arabic{equation}}
\renewcommand{\thefigure}{S\arabic{figure}}
\renewcommand{\thetable}{S\arabic{table}}

\preprint{}

\title{Supplemental material for\\ ''Realization of nodal ring semimetal in pressurized black phosphorus''}

\author{Kazuto~Akiba}
\email{akb@okayama-u.ac.jp}
\affiliation{
Graduate School of Natural Science and Technology,
Okayama University, Okayama 700-8530, Japan
}

\author{Yuichi~Akahama}
\affiliation{
Graduate School of Science, University of Hyogo, Kamigori, Hyogo 678-1297, Japan
}

\author{Masashi~Tokunaga}
\affiliation{
The Institute for Solid State Physics, The University of Tokyo,
Chiba 277-8581, Japan
}

\author{Tatsuo~C.~Kobayashi}
\affiliation{
Graduate School of Natural Science and Technology,
Okayama University, Okayama 700-8530, Japan
}

\date{\today}
\begin{abstract}
Supplemental material includes Figs. S1-S8, Table S1, and Supplementary notes I--IV.
\end{abstract}

\maketitle

\tableofcontents	

\clearpage


\section{
Methods
}
\subsection{
Experiments
}
\begin{figure}[]
\centering
\includegraphics[]{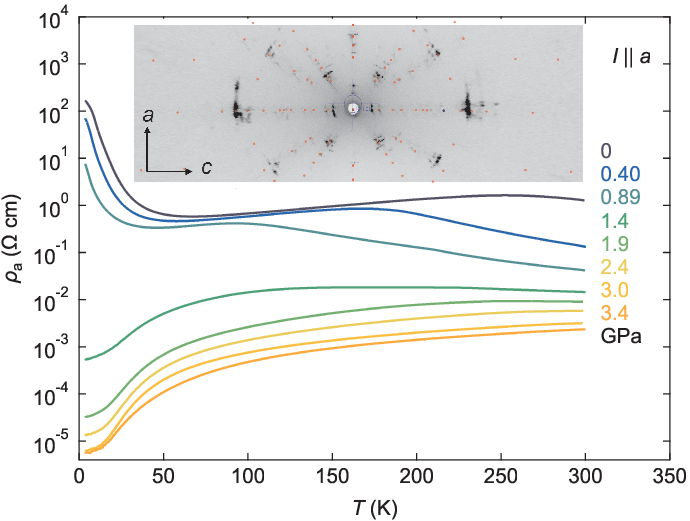}
\caption{
Temperature dependence of resistivity $\rho_{a}$ at various pressures up to 3.4 GPa.
The inset shows the Laue backscattering pattern of a sample.
The red dots represent the simulation based on Ref. \cite{Akahama_2020}.
The simulation was performed by QLaue \cite{qlaue_url}.
\label{figS01}}
\end{figure}

\begin{figure}[]
\centering
\includegraphics[]{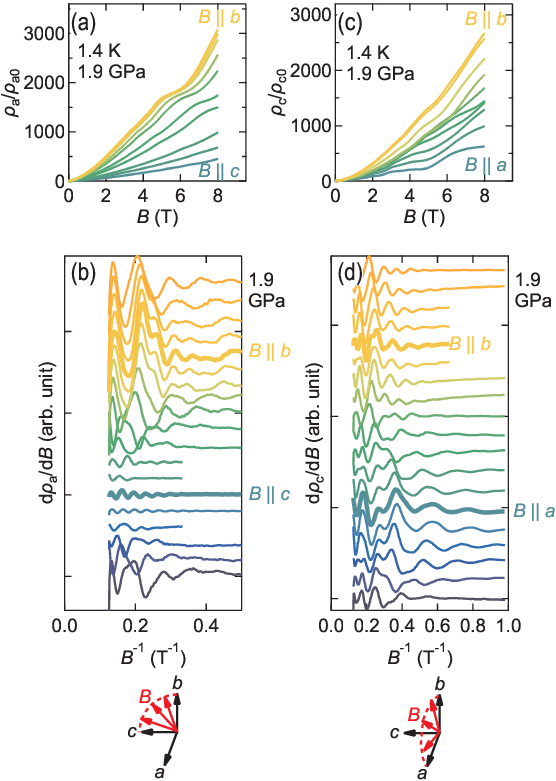}
\caption{
(a) Magnetic field dependence of resistivity $\rho_{a}$ normalized by zero-field value $\rho_{a0}$ and (b) $d\rho_{a}/dB$ at 1.4 K and 1.9 GPa.
Magnetic fields were rotated within the $bc$ plane and currents along the $a$ axis.
(c) Magnetic field dependence of resistivity $\rho_{c}$ normalized by zero-field value $\rho_{c0}$ and (d) $d\rho_{c}/dB$  at 1.4 K and 1.9 GPa.
Magnetic fields were rotated within the $ab$ plane and currents along the $c$ axis.
Data in (b) and (d) are vertically shifted for clarity.
\label{figS02}}
\end{figure}

\begin{figure}[]
\centering
\includegraphics[]{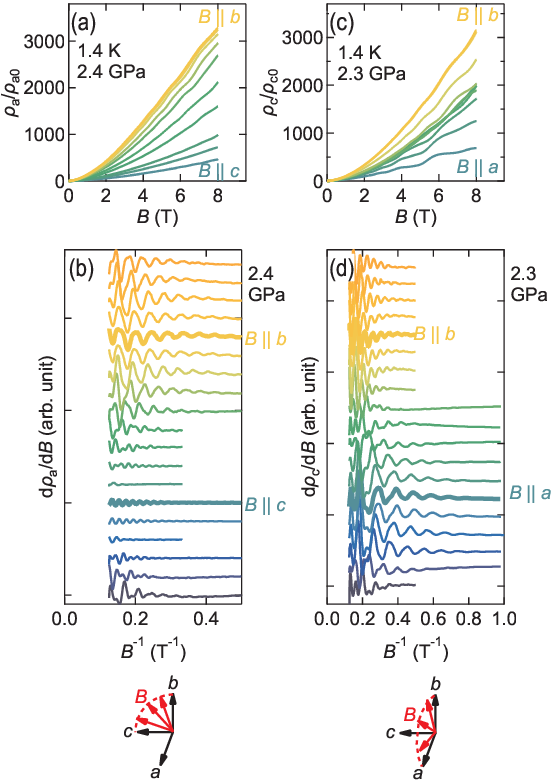}
\caption{
(a) Magnetic field dependence of resistivity $\rho_{a}$ normalized by zero-field value $\rho_{a0}$ and (b) $d\rho_{a}/dB$ at 1.4 K and 2.4 GPa.
Magnetic fields were rotated within the $bc$ plane and currents along the $a$ axis.
(c) Magnetic field dependence of resistivity $\rho_{c}$ normalized by zero-field value $\rho_{c0}$ and (d) $d\rho_{c}/dB$  at 1.4 K and 2.3 GPa.
Magnetic fields were rotated within the $ab$ plane and currents along the $c$ axis.
Data in (b) and (d) are vertically shifted for clarity.
\label{figS03}}
\end{figure}

\begin{figure}[]
\centering
\includegraphics[]{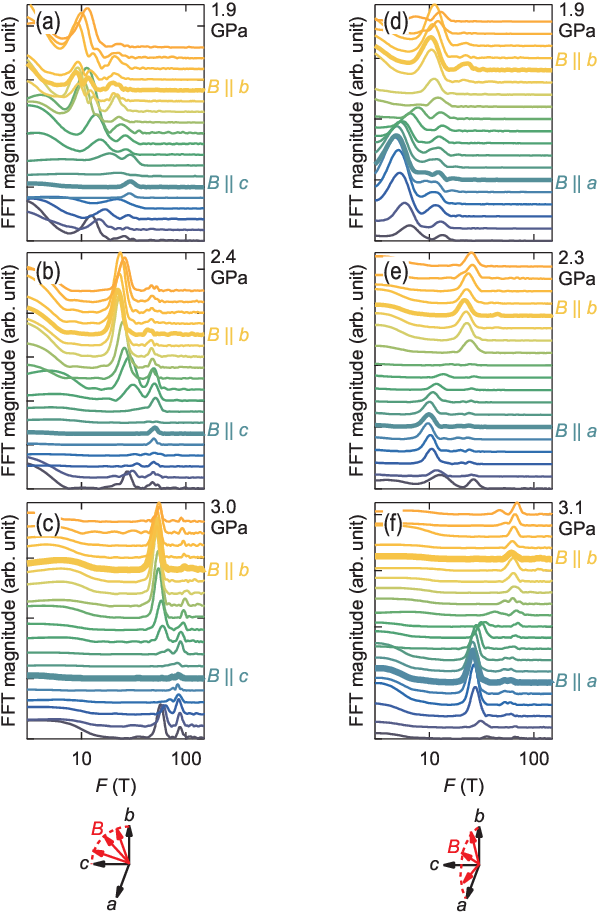}
\caption{
FFT spectra (a, b, c) under magnetic fields rotated within the $bc$ plane and (d, e, f) rotated within the $ab$ plane.
FFT analyses were performed on $d\rho_{a}/dB$ and $d\rho_{c}/dB$ at each pressure.
\label{figS04}}
\end{figure}

The resistivity measurements were performed using a standard four-probe method.
We used a Model 370 AC resistance bridge (Lake Shore Cryotronics, Inc.)
or a combined system of 2400 sourcemeter and 2182A nanovoltmeter (Keithley Instruments).
Measurements under zero-field conditions were performed using a He-gas-flow-type optical cryostat (Oxford Instruments) down to 2 K.
Measurements under magnetic fields of up to 8 T were performed using a superconducting magnet and variable-temperature insert (Oxford Instruments) down to 1.4 K.
The pressure in the sample space was determined from the superconducting transition temperature of Pb set near the sample \cite{Eiling_1981}.
The field-angular dependence of the resistivity under high pressure was measured using a homemade mechanical rotator that can uniaxially rotate the indenter-type pressure cell in the variable-temperature insert of a superconducting magnet.
Typical rotation steps adopted in the SdH measurements were 10\textdegree.
Fast Fourier transform (FFT) analyses were performed on $d\rho_{a}/dB$ and $d\rho_{c}/dB$ at each pressure.
We differentiated $\rho_{a}$ and $\rho_{c}$ with respect to $B$ and interpolated them at even intervals as a function of $B^{-1}$.
We obtained the FFT spectra from the $d\rho_{a}/dB$ and $d\rho_{c}/dB$ data with the application of the Hanning window function.
In Figs. \ref{figS01}, \ref{figS02}, \ref{figS03}, and \ref{figS04}, we show the temperature dependence of resistivity $\rho_{a}$ at various pressures, raw data of the Shubnikov de Haas oscillations at 1.9 GPa and 2.3-2.4 GPa, and all the FFT spectra used to construct the field angular dependence of $F$.

\subsection{
Calculations
}
Structural optimization and band structure calculation based on density-functional theory (DFT)
were performed using the Quantum ESPRESSO (QE) package \cite{Giannozzi_2009, Giannozzi_2017, Giannozzi_2020}.
We employed scalar-relativistic projector-augmented wave (PAW) pseudopotentials with the Perdew--Burke--Ernzerhof
(PBE) exchange correlation functional \cite{pseudo_wosoc}.
We used cutoffs of 70 Ry and 560 Ry for the plane-wave expansions of wave functions and charge density, respectively, and a $\Gamma$-shifted Monkhorst--Pack $12\times12\times12$ $\bm{k}$-point grid for the self-consistent calculation.
Self-consistent calculations were performed with a threshold of $1.0\times 10^{-8}$ Ry.
Structural optimization was performed using convergence thresholds of $1.0\times 10^{-5}$ Ry for the total energy change and $1.0\times 10^{-4}$ Ry/Bohr for the forces.
We constructed a tight-binding Hamiltonian using Wannier90 \cite{Pizzi_2020}.
We assumed 16 Wannier orbitals (P--$sp^3$ orbitals)
as initial projections to reproduce the DFT band structure at the Fermi level.
Visualization of the Wannier-interpolated band structure and FS were performed using WannierTools \cite{Wu_2017}
and FermiSurfer \cite{Kawamura_2019}.
Simulations of the quantum oscillation frequency $F$ and cyclotron effective mass were performed using the SKEAF code \cite{Rourke_2012}.
To express the small Fermi pockets with sufficient accuracy, the drawings of the pockets and calculations of $F$ were performed using a Wannier-interpolated Fermi surface with a dense $181\times181\times181$ $\bm{k}$-point mesh.

The simulation of the magnetoresistance effect was conducted based on the Boltzmann equation within the relaxation-time approximation using WannierTools \cite{Wu_2017, Zhang_2019}.
In this framework, the conductivity tensor is represented by
\begin{equation}
\dfrac{\sigma_{ij}^{(n)}(B\tau_n)}{\tau_n}=\dfrac{e^2}{4\pi^3}\int d\bm k v_i^{(n)}(\bm k) \bar{v}_j^{(n)}(\bm k, B\tau_n)\left(-\dfrac{\partial f_{FD}}{\partial \epsilon}\right)_{\epsilon=\epsilon_n(\bm k)}.
\label{eq_CF}
\end{equation}
Here, $e$, $f_{FD}$, and $n$ represent the elemental charge, Fermi--Dirac distribution function, and band index, respectively.
$\tau_n$ represents the relaxation time of the $n$th band, which is assumed to be independent of $\bm k$.
Because of the energy derivative of the Fermi--Dirac distribution function,
$\sigma_{ij}^{(n)}$ is determined by the states within the thermal energy width of $\sim k_B T$ near the Fermi level.
We set $T=15$ K to define the thermal energy width.
$\bm v^{(n)}(\bm k)$ represents the velocity defined by the gradient of energy in reciprocal space as follows:
\begin{equation}
\bm v^{(n)}(\bm k)=\dfrac{1}{\hbar}\dfrac{\partial \epsilon_n(\bm k)}{\partial \bm k}.
\end{equation}
$\bar{\bm v}^{(n)}(\bm k, B\tau_n)$ represents the weighted average of the velocity over the orbit, which is defined as
\begin{equation}
\bar{\bm v}^{(n)}(\bm k, B\tau_n)=\int_{-\infty}^{0}\dfrac{d(Bt)}{B\tau_n}e^{Bt/B\tau_n}\bm v^{(n)}[\bm k(t)].
\label{eq_wv}
\end{equation}
The historical motion of $\bm k(t)$ under a magnetic field $\bm B$ was obtained using the equation of motion
\begin{equation}
\dfrac{d\bm k(t)}{dt}=-\dfrac{e}{\hbar}\bm v^{(n)}[\bm k(t)]\times \bm B,
\end{equation}
where $\bm k(t=0)=\bm k$.
We adopted a $181^3$ $\bm k$ mesh for integration over the first Brillouin zone.
The total conductivity is obtained from $\sigma_{ij}/\tau=\sum_{n}\sigma_{ij}^{(n)}/\tau_n$,
assuming $\tau_n=\tau$ is independent of the band index.
The resistivity $\rho_{ij}\tau$ is obtained as the inverse matrix of $\sigma_{ij}/\tau$.

\clearpage

\section{
Structural optimization and the effect of the spin-orbit coupling
}
\label{app_fpc}

\begin{figure}[]
\centering
\includegraphics[]{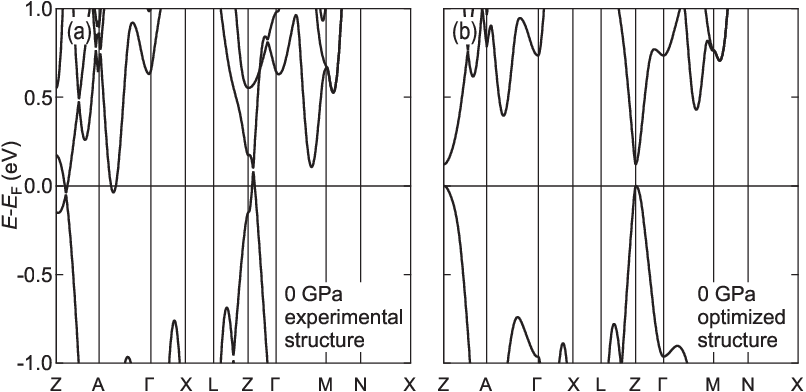}
\caption{
Energy band structures at ambient pressure based on (a) the experimental crystal structure reported in Ref. \cite{Akahama_2020}
and (b) the optimized crystal structure.
\label{figS05}}
\end{figure}

\begin{table}[]
\caption{\label{tab_opt}
Experimentally reported (expt., taken from Ref. \cite{Akahama_2020}) and fully optimized (opt.) crystal structures.
$a$, $b$, and $c$ represent the lattice constants, and $(0, y, z)$ is the atomic coordinate of 8f-P site.
$V=abc$ is the volume of the unit cell.
}
\begin{ruledtabular}
\begin{tabular}{lllllll}
&$a$ (\AA)& $b$ (\AA) & $c$ (\AA) & $y$ & $z$ & $V$ (\AA$^3$)\\
\hline expt. (0 GPa)&3.31400&10.4774&4.37536&0.10161&0.08074&151.921\\
opt. (0 GPa)&3.310&11.33&4.560&0.09379&0.08640&171.0\\
\hline expt. (3.18 GPa)&3.31429&10.0742&4.17685&0.10701&0.07337&139.460\\
opt. (1.5 GPa)&3.313&10.58&4.373&0.1015&0.07964&153.3\\
\end{tabular}
\end{ruledtabular}
\end{table}

First, we evaluated the computational conditions required to reproduce the semiconducting band structure at ambient pressure.
We performed a calculation using an experimental crystal structure reported in Ref. \cite{Akahama_2020}, which is in the first row of Tab. \ref{tab_opt}.
Contrary to the experimental fact, this results in a semimetallic band structure even at ambient pressure, as shown in Fig. \ref{figS05}(a).
The mismatch has been resolved using the fully optimized structure shown in the second row of Tab. \ref{tab_opt}, whose band structure is shown in Fig. \ref{figS05}(b).
Here, the Fermi level is taken at the top of the valence band.
Although the band gap (0.121 eV) at the $Z$ point is underestimated compared with the experimental value (0.3 eV), this is due to the potential difficulty associated with calculating the band structure of narrow gap semiconductors \cite{Qiao_2014}.
The band gap and optimized crystal structure obtained in this study agree with a previous calculation using the same type of pseudopotential and exchange-correlation functional \cite{Qiao_2014}.
Thus, we infer that the pressure-induced Lifshitz transition can be reasonably depicted using a fully optimized structure under hydrostatic compression.

The correspondence of the pressure between experiments and calculations is not straightforward because the starting structural parameters and the band gap in the calculations are different from the experimental values.
Since the band gap is underestimated at ambient pressure, we can assume that the DFT calculation tends to show the band structure at a pressure higher than the input value.
To make a reasonable comparison, we employed a calculation that has a volume ratio $V/V_0$ close to the experimental value.
Here, $V_0$ represents the volume at ambient pressure.
In the main text, we adopted an SdH result obtained at 3.0-3.1 GPa to determine the geometry of the Fermi surface.
Referring to the crystal structure at 3.18 GPa reported in a previous study (the third raw in Tab. \ref{tab_opt}), $V/V_0$ is assumed $\sim 0.92$.
Correspondingly, we adopted a computational result obtained using an optimized structure under hydrostatic compression corresponding to 1.5 GPa.
The fourth raw in Tab. \ref{tab_opt} represents the optimized structure.
The $V/V_0$ in this case is $\sim 0.90$, which is slightly smaller than the above value
but is assumed to be reasonable.
The agreement of $F$ between the experiment and calculation indicates the validity of the above comparison.
We confirmed that the conduction band on the $\Gamma$--$A$ path touches the Fermi level above 2.0 GPa ($V/V_0\sim 0.88$).
This situation requires another branch in the angular dependence of $F$, which can be excluded in the present case.

\begin{figure}[]
\centering
\includegraphics[]{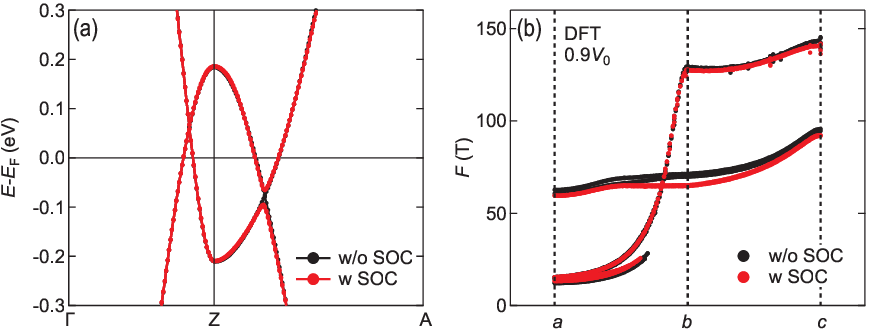}
\caption{
(a) Band structure along the $\Gamma$--$Z$--$A$ path and
(b) field angular dependence of the SdH frequency without (black) and with (red) spin-orbit coupling.
\label{figS06}}
\end{figure}

Here, we comment on the effect of spin-orbit coupling (SOC).
To test the effect of SOC, we calculated the band structure using the full-relativistic PAW pseudopotential with the PBE exchange-correlation functional \cite{pseudo_wsoc}, whose results are shown in Fig. \ref{figS06}(a).
Compared with the scalar-relativistic case, a small energy gap ($\sim 30$ meV) is discernible at the band-contact point on the $Z$--$A$ path, whereas the overall dispersion remains unchanged.
Additionally, the SOC causes only a slight change in the absolute value of the SdH oscillation frequency, as shown in Fig. \ref{figS06}(b).
Thus, we show the scalar-relativistic calculation in the main text.

\clearpage

\section{
Hall resistivity calculated based on the ellipsoidal Fermi surface model
}

\begin{figure}[]
\centering
\includegraphics[]{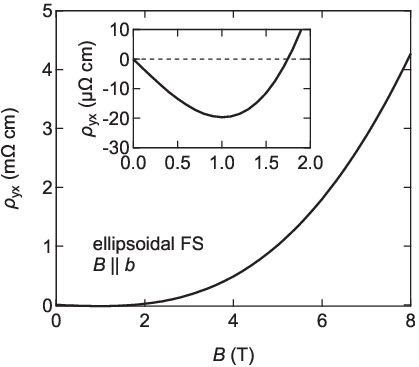}
\caption{
Magnetic field dependence of the in-plane ($a$-$c$ plane) Hall resistivity $\rho_{yx}$ with $B \parallel b$ calculated based on the ellipsoidal Fermi surface model.
The inset shows the magnified view in the low-field region below 2 T.
\label{figS07}}
\end{figure}

We calculated the in-plane Hall response based on the ellipsoidal Fermi surface model to check the quantitative validity.
Figure \ref{figS07} shows the magnetic field dependence of the in-plane ($a$-$c$ plane) Hall resistivity $\rho_{yx}$ with $B \parallel b$.
The inset in Fig. \ref{figS07} shows the magnified view in the low-field region below 2 T.
The sign of $\rho_{yx}$ is initially negative when $B$ is increased from 0 T, whereas the sign becomes positive between 1.5 and 2 T.
Similar sign inversion was observed in our previous experiment \cite{Akiba_2017}.
Referring to our previous measurements at 2.2 GPa and 2 K, $\rho_{yx}=6.7$ m$\Omega$ cm at 8 T and $|\rho_{yx}|=100$ $\mu\Omega$ cm at the local minimum.
Considering the tendency that these become smaller as the pressure increases, the agreement between the experiment (2.2 GPa) and model calculation (based on the Fermi surface at around 3 GPa) is reasonable.
The above comparison further supports that the obtained mobility and carrier density well explain the transport properties of pressurized BP.

\section{
Pressure-induced deformation of the hole pocket observed in the experiment
}
\begin{figure}[]
\centering
\includegraphics[]{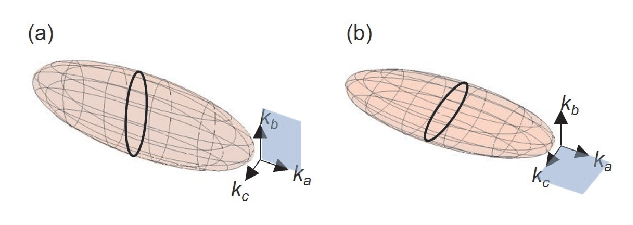}
\caption{
Schematic image of the hole pocket (a) at 1.9 GPa and (b) at 3.0-3.1 GPa.
Each pocket is approximated by a simple ellipsoid, and the anisotropy is exaggerated for clarity.
Each pocket has the largest cross-section when the magnetic field is perpendicular to the blue plane.
\label{figS08}}
\end{figure}

We show in Fig. \ref{figS08} the schematic image of the hole ($h$) pocket
expected from the experimental SdH oscillation.
In Fig. \ref{figS08}, the $h$ pocket is approximated by a simple ellipsoid, and the anisotropy is exaggerated for clarity compared to the real situation.
Continuous deformation from Fig. \ref{figS08}(a) to Fig. \ref{figS08}(b) by application of pressure
would explain the SdH results shown in the main text.

\bibliography{reference}